\documentclass[10pt,conference]{IEEEtran}

\usepackage{graphicx}

\graphicspath{{./figs/}}
\DeclareGraphicsExtensions{.eps}

\usepackage{longtable}
\usepackage{dcolumn}
\usepackage{multicol,multirow}
\usepackage{booktabs}
\usepackage{tabularx}
\usepackage{array}

\usepackage{hyphenat}

\usepackage[compress]{cite}

\usepackage{paralist}
\usepackage{enumerate}

\usepackage{subfig}

\usepackage{balance}

\usepackage{url}

\usepackage{amstext}
\usepackage{amsopn}

\usepackage[cmex10]{amsmath}
\usepackage{amssymb}
\usepackage{amsfonts}
\usepackage{amsthm}

\usepackage{verbatim}

\newcommand{\reffig}[1]{\mbox{Fig.~\ref{#1}}}

\newcommand{\refthm}[1]{\mbox{Thm.~\ref{#1}}}

\newcommand{\reflem}[1]{\mbox{Lem.~\ref{#1}}}

\newcommand{\refcor}[1]{\mbox{Cor.~\ref{#1}}}

\newcommand{\refasm}[1]{\mbox{Assm.~\ref{#1}}}

\newcommand{\refdef}[1]{\mbox{Def.~\ref{#1}}}

\newcommand{\refsec}[1]{\mbox{Sec.~\ref{#1}}}
\newcommand{\refSec}[1]{\mbox{Section~\ref{#1}}}

\newcommand{\refeqn}[1]{\mbox{\eqref{#1}}}

\renewcommand{\Pr}[1]{\ensuremath{\mathrm{Pr[}{#1}{\mathrm{]}}}}

\newcommand{\Ex}[1]{\ensuremath{\mathrm{E[}{#1}\mathrm{]}}}

\DeclareMathOperator{\sdom}{\ensuremath{\succ}}

\DeclareMathOperator{\mono}{{\scriptstyle\overset{\sdom}{\nearrow}}}

\usepackage{xspace}
\newcommand\abbrstyle{}
\newcommand{\etal}{\abbrstyle{et~al}.\@\xspace}
\newcommand{\ie}{\abbrstyle{i.e.}\xspace,\xspace}
\newcommand{\eg}{\abbrstyle{e.g.}\xspace,\xspace}
\newcommand{\etc}{\abbrstyle{etc}.\@\xspace}
\newcommand{\vs}{\abbrstyle{vs.}\@\xspace}
\newcommand{\cf}{\abbrstyle{cf.}\@\xspace}
\newcommand{\rv}{\abbrstyle{r.v.}\@\xspace}
\newcommand{\rvs}{\abbrstyle{r.v.s}\@\xspace}

\newtheorem{lemma}{Lemma}
\newtheorem{corollary}{Corollary}
\newtheorem{theorem}{Theorem}
\newtheorem{assumption}{Assumption}

\newtheorem{definition}{Definition}

\newtheorem{conjecture}{Conjecture}

\usepackage[nice]{nicefrac}

\newif\ifderivation\derivationfalse
\newif\ifplots\plotsfalse

\newcommand\ete{\emph{SrcFwd}\xspace}

\newcommand\hbh{\emph{IntFwd}\xspace}

\begin{document}

\newcommand\figwidth{0.55}
\newcommand\colfigwidth{0.85}
\newcommand\captionbefore{-0.70in}
\newcommand\captionafter{-0.2in}

\newcommand\subsectioninline[1]{\subsection{{#1}}}

%
%

\title{On Leveraging Partial Paths \\
in Partially-Connected Networks}

\author{%
\IEEEauthorblockN{%
Simon Heimlicher, 
Merkouris Karaliopoulos, 
Hanoch Levy\IEEEauthorrefmark{1} %
and
Thrasyvoulos Spyropoulos
}
\IEEEauthorblockA{%
Computer Engineering and Networks Laboratory\\
ETH Zurich, Switzerland
}
\IEEEauthorblockA{%
\IEEEauthorrefmark{1}
\footnotesize{Hanoch Levy is on leave absence from the School of
Computer Science, Tel Aviv University, Israel}
}
}

\maketitle


\begin{abstract}

Mobile wireless network research focuses on scenarios at the extremes of 
the network connectivity continuum where the probability of all nodes 
being connected is either close to unity, assuming connected paths 
between all nodes (mobile ad~hoc networks), or it is close to zero, 
assuming no multi-hop paths exist at all (delay-tolerant networks). In 
this paper, we argue that a sizable fraction of networks lies between 
these extremes and is characterized by the existence of \emph{partial 
paths}, \ie multi-hop path segments that allow forwarding data closer to 
the destination even when no end-to-end path is available.
A fundamental issue in such networks is dealing with disruptions of 
end-to-end paths. Under a stochastic model, we compare the performance 
of the established end-to-end retransmission (ignoring partial paths), 
against a forwarding mechanism that leverages partial paths to forward 
data closer to the destination even during disruption periods. Perhaps 
surprisingly, the alternative mechanism is not necessarily superior. 
However, under a stochastic monotonicity condition between current \vs 
future path length, which we demonstrate to hold in typical network 
models, we manage to prove superiority of the alternative mechanism in 
stochastic dominance terms.
We believe that this study could serve as a foundation to design more 
efficient data transfer protocols for partially-connected networks, 
which could potentially help reducing the gap between applications that 
can be supported over disconnected networks and those requiring full  
connectivity.

\end{abstract}

\section{Introduction\label{sec:intro}}

More and more people nowadays carry a device with wireless networking 
capabilities. The majority of laptops, organizers, and other
portable devices provide wireless networking; whereas in the cellular 
phone market this feature is trickling down from smart phones to the 
mainstream.
Such devices traditionally need infrastructure networks to communicate 
with each other, using protocol optimizations
of the TCP/IP suite to make up for the additional challenges of
wireless environments~\cite{tcp3g}. Nevertheless, their increasing
ubiquity creates opportunities for networking such
devices ``on the fly'' or in ``ad hoc'' mode, bypassing or
extending infrastructure, for applications ranging from social
networks to multi-player gaming.

However, operating such a network of mobile nodes in ad~hoc mode
(traditionally called a \emph{MANET}, for Mobile Ad Hoc Network)
presents a number of challenges for transport and routing
protocols. Initially, it was commonly assumed that MANETs are
always connected, i.e. \emph{each node has an end-to-end path to
every other node with probability one}; various mechanisms were
proposed to discover and maintain such paths, and traditional
transport mechanisms (or modifications) were assumed to provide
support for all known applications. Yet, frequent path changes
resulting from node mobility, wireless propagation effects, nodes
powering down, etc. induce a significant overhead and performance
penalty for these protocols as opposed to the infrastructure-based
counterparts. What is more, it has been recently recognized
that, when network density becomes lower, these networks
experience frequent disconnections, implying that no end-to-end
paths exist most of the time, or \emph{connectivity occurs with
probability zero}. At this end of the connectivity continuum, nodes
are assumed to be relatively isolated; in occasional node
encounters (``contacts''), forwarding decisions are made 
\emph{speculatively}, trying to predict future contacts with the
destination (\eg based on mobility patterns, social
relationships, \etc), but with no knowledge about end-to-end
paths. As a result, only few asynchronous applications with high
tolerance in delay can be supported under this network model, commonly 
referred to as \emph{DTN} (Delay Tolerant Networking).

In this work we argue that there is a sizable region in between the two 
extremes of the network connectivity continuum, where the network is not 
connected (and thus MANET protocols suffer), yet more optimistic 
assumptions are in order than commonly made in DTNs. In a 
\emph{partially-connected network}, nodes may not always be able to 
reach all other nodes, still they can reach a subset of them. If this 
subset is sufficiently variable over time, \eg due to node mobility, 
then there will be multi-hop path segments that can be used to forward 
data closer to the destination node, even during periods where the full 
path is disrupted. We refer to such path segments as \emph{partial 
paths}.

Partial paths correspond to additional transmission opportunities
that can be exploited by known network data forwarding mechanisms
such as hop-by-hop transport protocols
\cite{bibYSHbhbCongestionWirelessMultihop,bibHBMPTransportLayerRevisited} 
and route salvaging
mechanisms (for example, \cite{slr}) proposed in the context of
MANET routing, in order to achieve more efficient data transport
over these networks. This is in stark contrast to most algorithms 
proposed for DTNs \cite{jain04} that only exploit
single-hop paths at a time. Specifically, given a path breaks
after the message has traversed a partial segment of it, the
crucial question we are interested in answering is whether to try
forwarding data from the intermediate node before the breakage
\emph{(intermediate forwarding)} or from the source node
\emph{(source forwarding)}. Whereas the superiority of the former
might be ``intuitively'' clear in a network with fixed topology, it is
not obvious in a network whose topology varies stochastically due
to frequent link failures and node movements. Our work, therefore,
intentionally focuses on the concept of partial paths and these
two basic mechanisms rather than on particular protocols.
Specifically, we argue for the existence of partial paths, and
\emph{analytically explore} what are the conditions under which
``intermediate forwarding'' or ``local recovery'' techniques like
the above would result in more efficient data forwarding in
partially-connected networks.

The main contributions of this paper are as follows. First, we
demonstrate in \refsec{sec:networks} that partial paths do exist
under a wide range of network conditions. Further, when a path
breaks at an intermediate point, either an alternate partial path
from this point towards the destination exists or the intermediate
hop will have to ``wait'' until a new partial path for the
remaining distance arises. In this case, performance will depend
on the alternate route length (hop count) statistics and its relation to 
the primary route length. Using the network model introduced in 
\refsec{sec:model}, we show in
\refsec{sec:counterex} that---counter to intuition---if the alternate 
route length is
positively correlated with the primary route length; or even if the 
expected value of the alternate route length is monotonically increasing 
in the primary route length, superiority does not necessarily hold.
Nonetheless, we are able to show that if the length of the alternate 
route is \emph{stochastically monotonic} in the length of the primary 
route, then the time to deliver a packet under source forwarding 
stochastically dominates that under intermediate forwarding. In 
\refsec{sec:superiority}, we
introduce the concepts of stochastic dominance and stochastic
monotonicity, which we then use to prove this claim.
In \refsec{sec:alternate-monotonicity} we provide strong
evidence that the alternate route length is indeed stochastically
monotonic in the primary route length---leading to the desired result of 
superiority. \refSec{sec:related} discusses related work, and 
\refsec{sec:conclusion} concludes the paper.


%
\section{Under Which Conditions\\
Are Networks Partially-Connected?\label{sec:networks}}

In this section, we shed more light into the existence of
``partially-connected'' networks. Specifically, we are interested
in identifying whether scenarios exist
where networks are disconnected (and thus MANET approaches
would not be applicable or efficient), but still have sufficiently
large connected components with partial paths of more than one
hops (that DTN approaches would ignore).

\subsectioninline{Partial Paths in Stochastic Network Models}
We have looked
into the network connectivity dynamics using Monte Carlo
simulations. We spread $N$ nodes uniformly over a toroidal area
$A$ and add links between pairs of them according to the geometric
link model, \ie a link between two nodes exists as long as their
distance is smaller than the node transmission range $r_0$. We vary $A$ 
to obtain the desired network density, expressed by
the expected node degree $d$ as
$ 
  \label{eqn:deg-density}
  d = \frac{(N-1)\cdot \pi r_0^2}{A}.
$ 
For each $\{N,A\}$ tuple, simulations are repeated $M=10^5$
times with different seeds, each time yielding a different
snapshot of node distribution in $A$.

It is well-known by percolation results~\cite{Dousse:Thesis}
that, for asymptotically large networks, connectivity exhibits a
sharp phase-transition with respect to node density. Either almost
all nodes are connected into a large cluster (\emph{connected}
regime) or all nodes are isolated into many much smaller clusters
(\emph{disconnected} regime). Nevertheless, for finite $N$ values,
we argue that this phase transition is less sharp. Specifically, a
non-negligible range of network density values exists where
sizable clusters are formed; each node can reach a non-negligible
subset of other nodes (but not all) using partial paths of
multiple connected hops. To demonstrate this, we estimate and plot
the following three quantities:
\begin{itemize}
    \item \emph{Connectivity probability, $C$}: it is the
    probability that each node can reach every other node via a
    connected path, estimated as the ratio of the connected
    topologies over the full set size $M$ of random topologies
    \cite{Santi2003}.
  \item \emph{Reachability, $R$}: For each random topology, it equals
    the fraction of connected node pairs \cite{Perur2006a}. 
    \footnote{Note that an identical metric
    had been proposed earlier under the name ``connectivity
    index'' in \cite{Tang2003}. Irrespective of its name, the metric 
    serves as a less pessimistic measure
    of the communication capabilities provided by a sparsely-connected
    ad~hoc network.}
    \item \emph{Shortest path length, $P$}: For each random topology, it 
      denotes the average length of the shortest path in
      number of hops, considering all connected node pairs.
\end{itemize}
The plotted values for $R$ and $P$ are their averages over all $M$
random topologies.

\begin{figure}[h!tb]
  \begin{center}
  \includegraphics[scale=0.35]{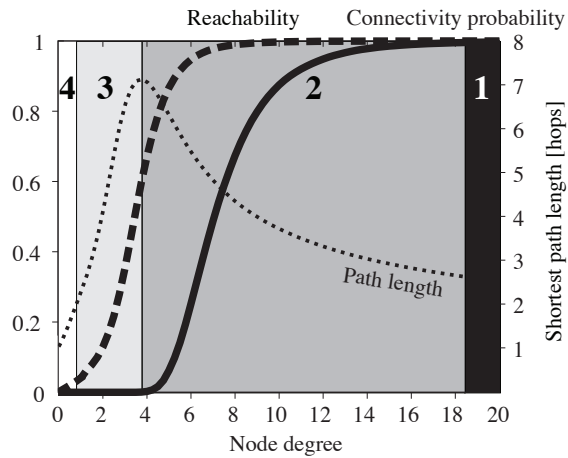}
  \caption{Partially-connected networks (grey areas) are situated 
  between connected (black) and disconnected networks (white). Model: 
  100 nodes uniformly distributed, geometric link model}
  \label{fig:conn-reach}
  \end{center}
\end{figure}

In \reffig{fig:conn-reach}, we show how $C, R$, and $P$ vary with
the node degree for
$N=100$ nodes. Observe that both the connectivity probability and
reachability exhibit a phase transition, albeit at different
values on the node degree axis. Whereas
the connectivity probability goes from zero to one
in the interval $d=4..20$, reachability increases more sharply in
$d=0..12$.

The sizable region of network density values where connectivity lies 
between zero and one represents the area of partially-connected 
networks, situated in between the (very) sparse networks commonly 
studied in the DTN community and the highly-connected networks studied 
in the MANET context. Note that the average length of the shortest path 
in the partially-connected region is greater than two; hence, the 
assumption that there are no multi-hop paths is too pessimistic even 
when $d=1$. Below, we identify four areas of interest in the 
connectivity continuum and explain whether and how intermediate 
forwarding can be applied in each:

\textbf{Area 1:} (Connectivity $\rightarrow 1$, Reachability
$\rightarrow 1$) This is the area (far right) commonly dealt with
by MANET research, where end-to-end paths are assumed to exist
most of the time. Yet, when the network is very dynamic (\eg high
node mobility), the initial path often breaks while the message is
en route. In this case, rather than dropping the message after it
has traversed a partial path, intermediate forwarding can re-route
it from that node on through an alternate path that the routing
protocol has cached or will discover after the break. This
mechanism is traditionally called \emph{salvaging} in the MANET
literature, and
is analyzed theoretically  in \refsec{sec:superiority}
(\refthm{thm:ts-dom-ti}).

\textbf{Area 2:} ($ 0 < $ Connectivity $ < 1$, Reachability $> 0$)
Nodes have end-to-end connectivity only for some fraction of
time. During this fraction, a tentative end-to-end routing path
between the source and destination nodes can be established, but
for the remainder of the time this path is disrupted
with only disconnected segments (referred to as partial paths) of it 
being up each time. Using
intermediate forwarding, nodes at the end points of these partial paths
can store en route data for the duration of the disruption and wait
until a new partial path comes up that allows to continue forwarding the 
data towards the destination. We compare the performance of such
intermediate forwarding against source forwarding analytically in
\refsec{sec:superiority} (\refthm{thm:ts-dom-ti-lat}).

\textbf{Area 3:} (Connectivity $\rightarrow 0$, Reachability $>
0$) The network is always partitioned. Nevertheless, node density
is high enough that each node can reach a non-negligible number of
other nodes within the same connected component (cluster) using
multi-hop partial paths. In this context, DTN algorithms are
usually applied or hybrid DTN/MANET schemes (MANET routing when
the destination is in the same cluster, and DTN routing if outside
the cluster)~\cite{Grossglauser:Secon06,Ott:DTN-MANET}. We believe that 
one could still take
advantage of existing partial paths using an appropriately
modified version of intermediate forwarding, and outperform
existing proposals. However, we defer this study for future work.

\textbf{Area 4:} (Connectivity $\rightarrow 0$, Reachability $
\rightarrow 0$) Nodes are essentially isolated. Forwarding
opportunities arise only when two nodes come in contact (\eg
through mobility), but multi-hop paths are rare. Such sparse
networks are properly treated by DTN schemes.

Summarizing, even for homogeneously distributed nodes, there is a
significant region of network density where a substantial portion
of nodes are connected by multi-hop paths even if the connectivity
probability is negligible. In the remainder of the paper, we will
show how this property allows one to perform significantly better
using local recovery methods to route around disruptions  instead of  
relying on end-to-end mechanisms.

\subsectioninline{Partial Paths in Real-life Scenarios}
Real-life networks are usually not uniform;
either the network area structure, or the node distribution in it,
or both may be non-uniform in many scenarios. Examples include
campus scenarios where nodes accumulate in areas of interest
(\eg library, cafeteria, classrooms) with less connectivity
available between these areas~\cite{Hsu:WiNMee06}, and vehicular networks
where nodes tend to gather at specific locations, \eg due to
 decelerating or stopping at junctions or traffic lights
~\cite{Grossglauser:Secon06}. This creates
concentrations of nodes (clusters) in specific network locations,
 thus yielding even more opportunities for multi-hop forwarding
than the uniform node distribution model predicts.

\subsectioninline{Discovering Partial Paths\label{sec:route-discovery}}
Given that partial paths exist, actually discovering and using them 
requires a routing algorithm. For source forwarding, we assume the use 
of some established routing protocol; however, in order to be able to 
use partial paths, intermediate forwarding requires a different routing 
approach. We believe that neither DTN nor MANET proposals can directly 
be applied for this task. We are currently working on a suitable 
algorithm and studying analytically its properties in terms of 
correctness, convergence, and performance. The key idea and main 
difference to existing protocols in the MANET area is to \emph{not} 
report route failures; instead, information about broken routes is still 
propagated. This allows nodes to keep forwarding data as far as possible 
towards the destination along those segments of broken routes that are 
still intact. Upon reaching the point of failure, packets are stored 
until a new---partial or complete---path becomes available, allowing 
further progress.

The selection among partial paths could be based on metrics such as 
those studied in the context of DTN routing, \eg age of last encounter 
\cite{Grossglauser:Secon06} or frequency of past encounters 
\cite{Burgess06}. But in contrast to DTN routing, these metrics would 
not only be used to discrimate among neighbors but to select among 
\emph{multi-hop} path segments.


%
\section{Modeling Source \vs Intermediate Forwarding\label{sec:model}}
\newcommand{\dotfigscale}{0.35}

We present our modeling assumptions for the network and the two
forwarding mechanisms under investigation; they are common to all
analytical arguments made in the remainder of this paper. As outlined in 
the previous section, we consider a scenario with a set of mobile nodes 
where partial  paths exist and can be discovered by a routing algorithm. 

\newcommand{\lsv}{\mathbf{L}}
\newcommand{\lse}{L}
\newcommand{\lsr}{l}
\newcommand{\lsvI}{\lsv^I}
\newcommand{\lseI}{\lse^I}
\newcommand{\lsrI}{\lsr^I}
\newcommand{\lsvS}{\lsv^S}
\newcommand{\lseS}{\lse^S}
\newcommand{\lsrS}{\lsr^S}
\newcommand{\lsl}{\alpha}
\newcommand{\lst}{\xi}
\newcommand{\lslI}{\lsl^I}
\newcommand{\lslS}{\lsl^S}

\begin{figure}[!t]
  \begin{center}
    \includegraphics[scale=0.70]{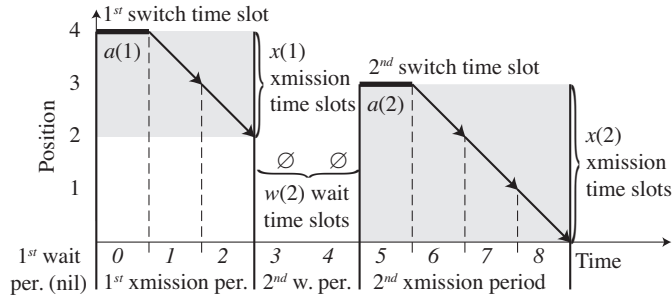}
    \caption{A packet begins $a(1)$ hops from the destination, but the 
    route breaks after $x(2)$ transmissions;  after $w(2)$ time slots, 
    an alternative route of length $a(2)$ hops is found}
    \label{fig:transmission-period}
  \end{center}
\end{figure}

\subsectioninline{Network Model\label{sec:network-model}}
In our model, time is slotted and we trace packets being transmitted 
from a source to a destination node. The random variable (\rv) $H(t)$ 
describes the hop count along the active or \emph{primary} route from 
the node holding the packet to the destination in time slot $t$. To 
refer to nodes that are \emph{according to the routing algorithm} on the 
primary route, we use the term \emph{position} $h$, meaning that this 
node has hop distance $h$ to the destination; the destination being at 
position $0$. Over time, the primary route may break and an 
\emph{alternate route} needs to be determined, which will then become 
the new primary route. These ``lifecycles'' of the route between source 
and destination node are numbered consecutively by $l=1,2,\dotsc$. The 
time required for a path to become available and the routing algorithm 
to establish a route is called \emph{waiting period} and represented by 
\rv $W(l)$ taking values $w(l)=0,1,\dotsc$. The subsequent 
\emph{transmission period} begins with a \emph{switch time slot}, in 
which the packet switches to position $a(l)=1,2,\dotsc$ corresponding to 
the length of the route. The switch time slot is followed by a sequence 
of $x(l)=0,1,\dotsc,a(l)$ \emph{transmission (``xmission'') time slots} 
during which the packet is transmitted towards the destination, as 
depicted in  \reffig{fig:transmission-period}.
Hence, in every transmission period $l$, the packet is transmitted from  
position $a(l)$ to positions $a(l)-1, a(l)-2,\dotsc,a(l)-x(l)$ and  
reaches the destination (position $0$) in the earliest transmission 
period where $a(l)-x(l)=0$. Every packet begins at initial hop distance 
$a(1)$; the initial wait time is $w(1)=0$.

Regarding the distributions of these \rvs, we assume that the link at 
the current position of a packet is up with probability $p$ or down with 
probability $1-p$, hence the distribution of \rv $X$ is geometric. For  
the waiting time $W$ and length of the alternate route $A$, we will 
assume that they depend only on the length of the primary route, \ie the 
position of the node requesting the alternate route. 
As will be described next, which node requests the alternate route is 
where the source and intermediate forwarding mechanisms differ.

\subsectioninline{Source and Intermediate 
Forwarding\label{sec:forwarding-model}}
Consider a packet that begins transmission period $l$ at some position 
$a(l)>0$ (the \emph{source node}), and is then
transmitted across $x(l)>0$ hops until it gets stuck at an  
\emph{intermediate node} at position $a(l)-x(l) > 0$. Now, under 
\emph{source forwarding (\ete)},
the intermediate node \emph{discards} the packet and the source node 
requests an alternative route to the destination; hence the distribution 
of $W(l+1)$ and $A(l+1)$ is conditioned on $a(l)$.
In contrast, under \emph{intermediate forwarding (\hbh)},
the intermediate node \emph{stores} the packet and requests
an alternative \emph{partial route} from itself to the destination; thus 
the distribution of $W(l+1)$ and $A(l+1)$ is conditioned on $a(l)-x(l)$.
This is the only difference between the two forwarding mechanisms, hence  
the relationship between the length of the primary and the alternate 
route determines the advantage of intermediate forwarding.


%
\section{IntFwd \emph{Not Necessarily} Superior\label{sec:counterex}}

In our model, source and intermediate forwarding differ as to which node 
continues forwarding a packet when the primary route fails; in 
particular, the length of the alternate route (hop count) depends on the 
length of the primary route from the requesting node to the destination. 
In this section, we use \rv $H$ to describe the length of the primary 
route, $A$ for the alternative route, and $A_h$ to describe the length 
of the alternate route conditioned on the primary route having length 
$h$. We are interested in a sufficient condition on this relationship to 
guarantee that intermediate is faster than source forwarding. Such a 
condition must guarantee that the time to forward a packet to the 
destination along a route of given length $h$, denoted by $T_h$, is 
monotonically increasing in $h$, \ie $T_k \ge T_j$ holds for every 
$k>j$.

Contrary to intuition, even if the length of the alternate route is 
tightly related to that of the primary route, monotonicity does not 
necessarily hold. In particular, we show that neither the positive 
correlation between alternate and primary route length 
(\refthm{thm:mono-corr}) nor the monotonic increase of the 
\emph{expected} alternate route length in the primary route length 
(\refthm{thm:mono-exp}), are sufficient conditions.

\begin{figure}[h!tb]
  \begin{center}
    \includegraphics[scale=\dotfigscale]{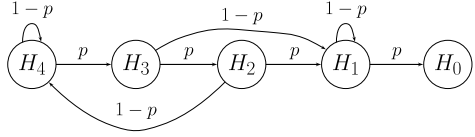}
  \end{center}
  \caption{Counter example for positive correlation between alternate 
  and primary route length}
  \label{fig:counter-ex-corr}
\end{figure}
\begin{theorem}\label{thm:mono-corr}
  Assume that the length of the alternate route is positively correlated
  with the length of the primary route, \ie the correlation
  coefficient of \rvs $H$ and $A$, $\rho_{HA} > 0$. Then, $T_k \ge T_j$ 
  does not necessarily hold for every $k > j$.
\end{theorem}
\begin{IEEEproof}
  By counter example. Let the
  primary route length take values $1$ to $4$ with equal probability.
  Assume that the corresponding alternate route lengths $A_h$ are 
  related to the primary route lengths through the following conditional 
  distribution:
  $A_1=1, A_2=4, A_3=1, \text{ and } A_4=4$, all with probability
  one (\cf \reffig{fig:counter-ex-corr}).

  Observe that alternate and primary route length are positively
  correlated with correlation coefficient $\rho_{HA}
  = \nicefrac{1}{\sqrt{5}}~>~0$. To derive this, we used
  $\rho_{HA}= \nicefrac{\sigma_{HA}}{(\sigma_H \sigma_A)}$,
  where $\sigma_{HA} = \sum_k\sum_l\left((h_k-\mu_H)(a_l-\mu_A)
  \Pr{H=h_k, A=a_l}\right)$.

  Despite the positive correlation, the alternate route of $H_2$ is four hops, whereas the one of $H_3$ is only one hop. The expected values of
  the time to destination $T_k$ coincide
  with the mean times to absorption starting from position
  $k$, when viewing the node chain as an absorbing Markov chain
  with a single absorbing state (destination node); \cf
  \cite{heimlicher2008} for the complete derivation. Using this, we can 
  relate $T_{3}$ and $T_{2}$ with the following equation:
  \begin{equation*}
    T_2 - T_3 = \frac{1+2p}{p^2} - \frac{2p^2 + 2}{p^3 - p^2 + p}.
  \end{equation*}
  This implies that $T_3 < T_2$ for all values of $p \in
  (0,\nicefrac{(\sqrt{5}-1)}{2})$; in other words, $T_{h}$ is not monotonically increasing in $h$.
\end{IEEEproof}

Next we show that even the \emph{expected} alternate route length being 
monotonically increasing in the primary route length does not guarantee 
that $T_{h}$ is monotonically increasing in $h$.

\begin{figure}[h!tb]
  \begin{center}
    \includegraphics[scale=\dotfigscale]{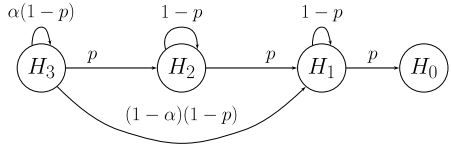}
  \end{center}
  \caption{Counter example for expected value of alternate route length 
  monotonically increasing in primary route length}
  \label{fig:counter-ex-expected}
\end{figure}
\begin{theorem}\label{thm:mono-exp}
  Assume $\Ex{A_k} \ge \Ex{A_j}$ for every $k > j$. Then $T_k \ge
  T_j$ does not necessarily hold for every $k > j$.
\end{theorem}
\begin{IEEEproof}
  By counter example. Let the primary route length take values
  $1,2$, or $3$; the alternate route lengths are $A_1=1,
  A_2=2$, each with probability one, whereas $A_3 = 3$ with
  probability $\alpha$ and $A_3 = 1$ with probability $1-\alpha$ (\cf 
  \reffig{fig:counter-ex-expected}).

  Observe that the expected length of the alternate route is
  monotonically increasing in the length of the primary route as long as $\nicefrac{1}{2}<\alpha
  < 1$: $\Ex{A_k} = 1,2,2\alpha+1, \text{ for } k=1,2,3$,
  respectively. Yet a packet at $H_3$ may cut through to $H_1$
  with non-negligible probability $(1-\alpha)(1-p)$.
  Invoking again the absorbing Markov chain argument, as in 
  \refthm{thm:mono-exp}, it can be shown (\cf
  \cite{heimlicher2008}) that:
  \begin{equation*}
    T_2-T_3 = \frac {2}{p}-\frac{3}{1-\alpha+\alpha p}.
  \end{equation*}
  implying that $T_3 < T_2$ for all values of $p \in
  (0,\nicefrac{(2-2\alpha)}{(3-2\alpha}))$.
  Consider, \eg $\alpha=\nicefrac{5}{8}$ and $p=\nicefrac{1}{3}$, then 
  $T_k = 5,10,6$ for $k=1,2,3$, which is not monotonically
  increasing in $k$.
\end{IEEEproof}


%
\section{IntFwd Superior Given\\
Stochastic Monotonicity of Route Length\label{sec:superiority}}

We now turn to a condition, \emph{stochastic monotonicity}, which as we  
will show yields that the time to deliver a packet to the destination 
under source forwarding \emph{stochastically dominates} that under 
intermediate forwarding. We first introduce the concept of stochastic 
monotonicity and derive a fundamental result on its transitivity; we 
then use this result to prove the above claim, first under the 
assumption of immediate availability of alternate routes, then under a 
relaxed assumption.
\begin{definition}[\emph{Stochastic Dominance --- Stochastic 
  Monotonicity}] \label{def:stoch-monotonicity} An \rv $X$ 
  \emph{stochastically
dominates} \rv $Y$, written as $X \succ Y$, if $\Pr{X > t} \ge
\Pr{Y > t} \; \forall\;  t$, or equivalently if
$
    F_X(t) \le F_Y(t)\;  \forall\;  t.
$
\footnote{This is also referred to as  \emph{First-order Stochastic 
Dominance}, \eg
\cite{Levy2006}.}
Drawing on the stochastic dominance concept, we define stochastic
monotonicity. An \rv $X$ is \emph{stochastically
monotonic} in \rv $A$, written as $X \mono A$, if $\Pr{X \ge x |
A=a_1} \ge \Pr{X \ge x | A=a_0} \; \forall \; a_1 > a_0
\label{eqn:x-mono-a}$; or if
$
  F_{X|A}(x|A=a_1) \le F_{X|A}(x|A=a_0)\; \forall\; a_1 > a_0.
$
\end{definition}

Stochastic monotonicity is \emph{transitive}
under the conditions outlined in \reflem{lem:transitivity}.
\begin{lemma}\label{lem:transitivity}
  Let $X, Y, Z$ be \rvs with strictly monotonic and
  continuously differentiable cumulative distribution functions (CDF).
  Then $Y \mono X$ implies $Z \mono X$ if the following holds:
  \begin{multline*}
    F_{Z|Y,X}(z|Y=y_1,X=x_1) \le F_{Z|Y,X}(z|Y=y_0,X=x_0) \\
    \; \forall \; y_1 > y_0, x_1 > x_0. \label{eqn:z-mono-xy}
  \end{multline*}
\end{lemma}
The proof is based on the law of total probability and provided in 
\cite{heimlicher2008}. In the following, we are mainly
interested in the corollary below, emerging from 
\reflem{lem:transitivity} as the special case where \rv $Z$ only
depends on \rv $X$ via $Y$.
\begin{corollary}
  Let $X, Y, Z$ be \rvs with strictly monotonic and
  continuously differentiable CDFs. If $Y \mono X$, and $Z \mono Y,$
  then $Z \mono X$ if $F_{Z|Y,X} = F_{Z|Y}$.
  \label{cor:transitivity}
\end{corollary}

We now use these concepts to compare source and intermediate forwarding, 
based on the network model and notation from \refsec{sec:model} and 
under the following assumption.
\begin{assumption}[Alternate route length]\label{asm:alternate}
  The distribution of the alternate route length $A$ is time-invariant 
  and is stochastically monotonic in the primary route length $H$, \ie 
  $A\mono H$ or equivalently if $F_A$ denotes the CDF of $A$, then 
  $F_{A|H}(a|H=k) \le F_{A|H}(a|H=j) \text{ for every } k > j$.
\end{assumption}
Evidence that this assumption holds in a sizable subset of networks is 
provided in \refsec{sec:alternate-monotonicity}.

\subsectioninline{Immediate Route Discovery}
In this part, we assume that alternate routes can be determined 
immediately, corresponding to Area 1 (\refsec{sec:networks}); this 
assumption is relaxed in \refsec{sec:alternate-lat}.
\begin{assumption}[Immediate route discovery]\label{asm:alternate-imm}
  The alternate route is available immediately, allowing the packet to 
  continue along the alternate route in the subsequent time slot.
\end{assumption}
We first show that the position of a packet forwarded under \ete 
stochastically dominates that of a packet under \hbh 
(\refthm{thm:hs-dom-hi}) if they start from the same initial position; 
this will imply that the same relationship holds for the time to deliver 
a packet to the destination (\refthm{thm:ts-dom-ti}). To this end, we 
will compare the positions of packet $S$ forwarded by $\ete$ against 
packet $I$ forwarded by \hbh over the course of time; when referring to 
one of them, we will use corresponding superscripts $S$ and $I$ as 
appropriate. 

\begin{figure}[!t]
  \begin{center}
    \includegraphics[scale=0.70]{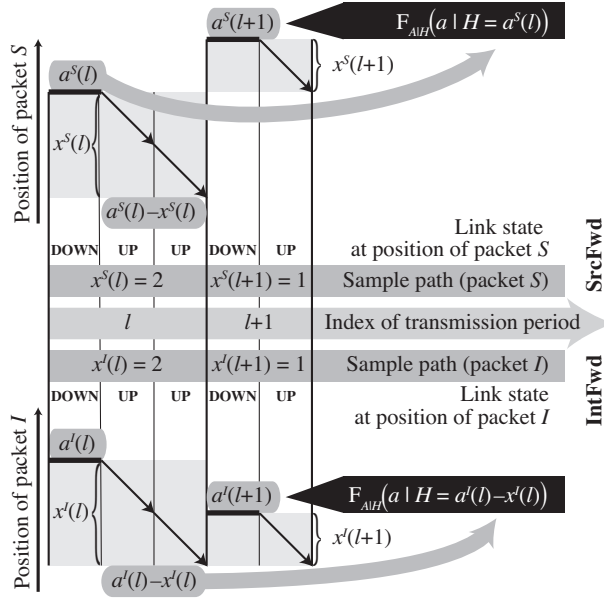}
  \end{center}
  \caption{Comparing the positions of packets $S$ and $I$ during 
  exemplary transmission periods $l$ and $l+1$ given they trace 
  identical link state sample paths}
    \label{fig:length-vs-position}
\end{figure}

For the purposes of the proof, we will track the two packets given the 
condition that the links at their positions are always in the same 
state, \ie up or down. This condition means that \rvs $X^S(l), X^I(l)$ 
yield \emph{identical outcomes} (``samples'') $x^S(l)=x^I(l)$ for every 
transmission period $l$; we refer to such a sequence of outcomes as a 
\emph{sample path} $\lsv$. As illustrated in 
\reffig{fig:length-vs-position}, in light of \refasm{asm:alternate-imm} 
the waiting periods shown in \reffig{fig:transmission-period} are nil 
and transmission periods back-to-back. A sample path describing the link 
states of $\tau$ time slots comprises outcomes 
$x(1),x(2),\dotsc,x(\lsl)$, where $\lsl$ is the number of switch time 
slots (\ie link down); similarly $\lst:=\tau - \lsl$ denotes the number 
of transmission time slots (link up).

For a finite number of time slots $\tau$, the number of possible link 
state sample paths is given by $2^\tau$ and can be enumerated; we use 
subscript $k$ to refer to individual sample paths $\lsv_k, 
k=1,2,\dotsc,2^\tau$ and corresponding quantities $\lsl_k, \lst_k$.
We say that a packet \emph{traces} a certain sample path $\lsv_k$ if the 
link at the packet's position is always in the state described by the 
sample path. The probability of this event is given by $\Pr{\lsv} = 
p^{\lst}\cdot (1-p)^{\lsl}$, as all links are independent and up with 
probability $p$ (\cf \refsec{sec:network-model}). In the following 
theorem, we will \emph{match corresponding sample paths} 
$\lsvS_k=\lsvI_k=\lsv_k$ and show that the claim holds under this 
condition; at the end we will un-condition this result using the law of 
total probability \cite{Durrett:prob}.
\begin{theorem}
  \label{thm:hs-dom-hi}
  Let packet $S$ be forwarded by \ete and packet $I$ by \hbh.
  If the positions of packets $S$ and $I$ at time $0$ are given by
  $a(1)=1,2,\dotsc$, then their future positions obey
  \begin{equation}
    H^S(\tau)  \sdom H^I(\tau) \quad \forall \; \tau\ge 1. 
    \label{eqn:hs-dom-hi}
  \end{equation}
\end{theorem}
\begin{IEEEproof}
  Let both packets $S$ and $I$ trace identical sample paths, \ie  
  $\lsvS_k=\lsvI_k=\lsv_k$; given this condition, denote their position 
  at time $t$ by $H^S_k(t), H^I_k(t)$ and their position in the 
  $l^{\text{th}}$ switch time slot by $A^S_k(l), A^I_k(l)$, 
  respectively. We use $\lsv_k[u]$ to denote a sample path prefix 
  comprising only the first $u \le \lsl_k$ elements 
  $x_k(1),x_k(2),\dotsc,x_k(u)$ of $\lsv_k$. We next use induction on 
  the length $u$ of identical sample path prefixes 
  $\lsv^S_k[u]=\lsv^I_k[u]=\lsv_k[u]$ to show that the following holds 
  for every length $u=1,2,\dotsc,\lsl_k$:
  \begin{align}
    A^S_k(u) &\sdom A^I_k(u). \label{eqn:as-dom-ai}
  \end{align}
  \emph{Inductive basis. }
  For $u=1$, we have $A^S_k(1)=A^I_k(1)=a(1)$. \\
  \emph{Inductive step. }
  Assume \refeqn{eqn:as-dom-ai} holds for $u$ and prove for $u+1$.
  By \refasm{asm:alternate}, the alternative route length $A(u+1)$ is 
  stochastically monotonic in some primary route length. For packet $S$,  
  this means that $A^S_k(u+1) \mono A^S_k(u)$, $A^S_k(u)$ being the 
  position the packet had in the previous switch time slot. For packet 
  $I$, it means that $A^I_k(u+1) \mono \bigl(A^I_k(u)-X^I_k(u)\bigr)$, 
  $A^I_k(u)-X^I_k(u)$ being the position the packet reached at the end 
  of the previous transmission period. Hence, the inductive assumption   
  $A^S_k(u)\sdom A^I_k(u)$ implies immediately $A^S_k(u) \sdom 
  \bigl(A^I_k(u)-X^I_k(u)\bigr)$, \ie stochastic dominance between the 
  primary route lengths in which the alternate route lengths are 
  stochastically monotonic; hence application of 
  \refcor{cor:transitivity} yields that \refeqn{eqn:as-dom-ai} also 
  holds for $u+1$.
  
  Still given that both packets trace identical sample paths,  
  \refeqn{eqn:as-dom-ai} valid for every $u=1,2,\dotsc,\lsl_k$ yields 
  that \refeqn{eqn:hs-dom-hi} holds \emph{for every switch time slot} of 
  these sample paths. In addition, their sample paths being identical 
  also means that both packets are transmitted simultaneously over the 
  same number of hops $x^S_k(u)=x^I_k(u)$ after every switch time slot,  
  hence \refeqn{eqn:hs-dom-hi} valid for switch time slots also yields 
  that it holds \emph{for every transmission time slot}.
  
  Summarizing, given that the packets trace identical sample paths 
  $\lsv^S_k=\lsv^I_k=\lsv_k$, their positions obey $H^S_k(t) \sdom 
  H^I_k(t)$, which can also be written as
  \begin{align}
    F_{H^S_k}(h,t)  &\le F_{H^I_k}(h,t).
    \label{eqn:hs-dom-hi-pik}
  \end{align}
  Summation over all possible sample paths $\lsv_k, k=1,2,\dotsc,2^\tau$ 
  then yields:
  \begin{align*}
    F_{H^S}(h,t) &= \sum_{k=1}^{2^\tau}\Pr{\lsv_k^S} F_{H^S_k}(h,t),
    \\
    F_{H^I}(h,t) &= \sum_{k=1}^{2^\tau}\Pr{\lsv_k^I} F_{H^I_k}(h,t).
  \end{align*}
  Since the sample paths are identical, they are also equiprobable 
  ($\Pr{\lsv^S_k}=\Pr{\lsv^I_k}$); this together with   
  \refeqn{eqn:hs-dom-hi-pik} implies $F_{H^S}(h,t) \le F_{H^I}(h,t)$, 
  which is equivalent to the claim.
\end{IEEEproof}

\newcommand{\fspe}{L}
\newcommand{\fspv}{\mathbf{L}}

In the following theorem, we show that the time to the destination of 
a packet forwarded by \ete stochastically dominates that of a packet 
forwarded by \hbh if they both start from the same position.
\begin{theorem}
  \label{thm:ts-dom-ti}
  Let packet $S$ be forwarded by \ete and packet $I$ by \hbh.
  If their positions at time $0$ are given by $a(1)$, then their times 
  to the destination obey
  \begin{equation*}
    T^S(\tau) \sdom  T^I(\tau) \quad \forall \; \tau \ge 1.
    \label{eqn:ts-dom-ti}
  \end{equation*}
\end{theorem}
\begin{IEEEproof}
  Observe, first, that since $0$ is the minimal outcome of \rv $H(t)$, 
  $\Pr{H(t)=0}=\Pr{H(t)\le 0}$. Second, the probability that a packet 
  reaches the destination by some time $t$ is the same as the 
  probability that, at time $t$, the packet is at position $H(t) = 0$, 
  \ie $\Pr{T \le t} = \Pr{H(t)=0}$.
  
  From \refthm{thm:hs-dom-hi}, $H^S(\tau) \sdom H^I(\tau)$ implies 
  $\Pr{H^S(\tau)\le 0} \le \Pr{H^I(\tau)\le 0}$, which by the first  
  observation is equivalent to $\Pr{H^S(\tau)= 0} \le \Pr{H^I(\tau)= 
  0}$. By the second observation, this implies $\Pr{T^S(\tau)\le \tau} 
  \le \Pr{T^I(\tau)\le \tau}\; \forall \; \tau \ge 1$, concluding the 
  proof.
\end{IEEEproof}
\subsectioninline{Accounting for Stochastic Route Discovery Latency\label{sec:alternate-lat}}
Lastly, we relax the assumption that the alternate route is always
available immediately. The analysis carried out so far assumes that an
alternate route is available immediately upon failure of the primary 
route. However, in many scenarios (\eg Area 2 in \refsec{sec:networks}) 
the intermediate (or source node) will have to wait for some time until 
an alternate route becomes available. This time mainly depends on the 
mobility patterns of the nodes and the network density.

In the following, we generalize our analysis by assuming that the time 
to discover an alternate route is an \rv $W$. We will make the 
following assumptions on $W$:
\begin{assumption}[Waiting time]\label{asm:alternate-latency}
  The distribution of $W$ is time-invariant and may depend only on the 
  length of the primary route $H$, similarly to how the length of the 
  alternate route $A$ depends on $H$ by  \refasm{asm:alternate}. 
  Specifically, $W$ is     stochastically monotonic in the length of the 
  primary route, \ie if $F_W$ denotes the CDF of $W$, then $F_{W|H}(w|H=k) 
  \le F_{W|H}(w|H=j) \text{ for every } k > j$.
\end{assumption}

The assumption of time-invariance holds when the mobility model is 
stationary and network density is time-invariant. Regarding stochastic 
monotonicity, we argue as follows: in the worst case, the intermediate 
hop will have to wait for a time longer than the \emph{mixing 
time}~\cite{Durrett:prob} for the given mobility model and network. In 
this case the waiting time to find an alternate route for the 
intermediate and source nodes follows the same distribution; such 
equality is allowed under our assumption of stochastic monotonicity. 
Nevertheless, it can be shown that, for many mobility models with high 
mixing time (\eg two-dimensional random walk) and sufficiently high node 
density, the waiting time till a new route is found from an intermediate 
node is actually smaller than for the source node.   

Making this assumption instead of \refasm{asm:alternate-imm}, we next 
claim that \refthm{thm:ts-dom-ti}
still holds.
\newcommand{\spv}{\mathbf{M}}
\newcommand{\spe}{M}
\newcommand{\spr}{m}
\newcommand{\spvI}{\spv^I}
\newcommand{\speI}{\spe^I}
\newcommand{\sprI}{\spr^I}
\newcommand{\spvS}{\spv^S}
\newcommand{\speS}{\spe^S}
\newcommand{\sprS}{\spr^S}
\newcommand{\spl}{\rho}
\newcommand{\splI}{\spl^I}
\newcommand{\splS}{\spl^S}
\newcommand{\ldv}{\mathbf{D}}
\newcommand{\lde}{D}
\newcommand{\ldl}{\Delta}
\begin{theorem}
  \label{thm:ts-dom-ti-lat}
  Under \refasm{asm:alternate-latency}, \refthm{thm:ts-dom-ti} holds, 
  namely
  \begin{equation*}
    T^S(\tau) \sdom  T^I(\tau) \quad \forall \; \tau \ge 1.
    \label{eqn:ts-dom-ti-lat}
  \end{equation*}
\end{theorem}
\begin{IEEEproof}
  Due to space constraints, we only give a sketch of the proof, which is  
  provided in \cite{heimlicher2008}. We again let packets $S$ and $I$ 
  trace pairs of corresponding sample paths and compare their times to 
  the destination under this condition. The difference to the previous 
  derivation is that the waiting time, previously zero, is now a random 
  variable. By \refasm{asm:alternate-latency}, the waiting time 
  following transmission period $l$, represented by \rv $W(l+1)$,
  depends on previous packet positions in the same way as the alternate 
  route length \rv $A(l+1)$; namely for \ete it depends on the position 
  of the source, $A^S(l)$, and for \hbh it depends on the position of 
  the intermediate node, $A^I(l)-X^I(l)$. This means that if we know the 
  exact sequence of positions a packet visits till it is delivered to 
  the destination, we also know the outcomes $a(l)$ and $x(l)$ in every 
  transmission sequence $l$, on which the subsequent waiting period's 
  duration, $W(l+1)$, depends.
  Therefore, for this theorem we use sample paths that include outcomes 
  $a(l)$ and $x(l)$ of every transmission period $l=1,2,\dotsc,\lsl$. 

  Let us ignore waiting periods for a moment. Using standard coupling 
  techniques \cite{Thorisson2000}, one can match sample paths of packets 
  $S$ and $I$ such that (i) they are equiprobable, (ii) $x^S(l)=x^I(l)$ 
  and (iii) $a^S(l) \ge a^I(l)$ is valid in every transmission period. 
  This can be achieved by defining all sample paths of packet $I$ such 
  that they end with the delivery of the packet, then match to each of 
  them a corresponding sample path of packet $S$ comprising the same 
  number of transmission periods such that properties (i)--(iii) hold. 

  Now, again considering waiting periods, observe that (ii) together 
  with (iii) implies directly $a^S(l) \ge a^I(l)-x^I(l)$. Since by  
  \refasm{asm:alternate-latency}, $W^S(l+1)\mono A^S(l)$ and 
  $W^I(l+1)\mono A^I(l)-X^I(l)$, this inequality implies $W^S(l+1)\sdom 
  W^I(l+1)$. Stochastic dominance is preserved under summation, hence 
  the total waiting time of every sample path pair also obeys this 
  relationship. This means that in addition to the (equal) number of 
  switch and transmission time slots, the time to the destination is 
  increased by the sum of all waiting periods; therefore, the total sum 
  of time slots (switch, transmission, and wait time slots) still obeys 
  the stochastic dominance relationship. Lastly, every sample path of 
  packet $I$ ending with the delivery of the packet and none of the 
  sample paths of packet $S$ delivering the packet earlier (if at all), 
  un-conditioning from the sample paths yields the claim.
\end{IEEEproof}


%
\section{Stochastic Monotonicity of Route 
Length\label{sec:alternate-monotonicity}}

Through the analysis in
\refsec{sec:counterex}-\ref{sec:superiority}, we have demonstrated
that the stochastic monotonicity of alternate path length (hop count) in
the primary path length is sufficient for intermediate forwarding
to outperform source forwarding. We now turn to the question of
whether this stochastic monotonicity indeed holds in realistic
wireless multi-hop networks. Whereas a rigorous proof along the
lines of the analysis in the previous sections appears hard to
devise, we draw on available analytical results in literature to
argue that such monotonicity exists.
\begin{figure*}[!htb]
  \begin{center}
       \subfloat[Distance given shortest path length]{
  \label{fig:dist-given-hops}
       \includegraphics[scale=0.35]{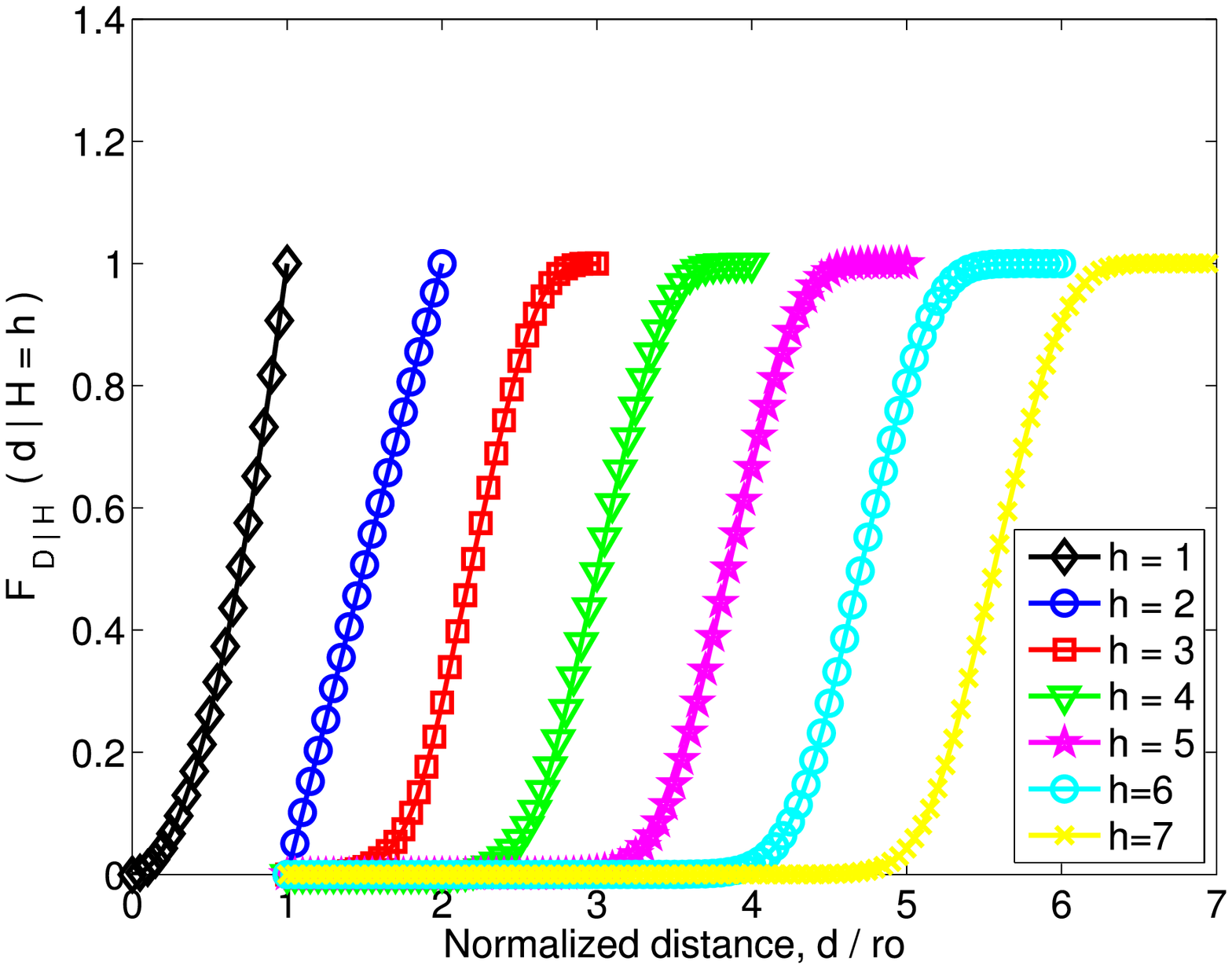}
       }
    \subfloat[Shortest path length given distance]
    {
    \includegraphics[scale=0.35]{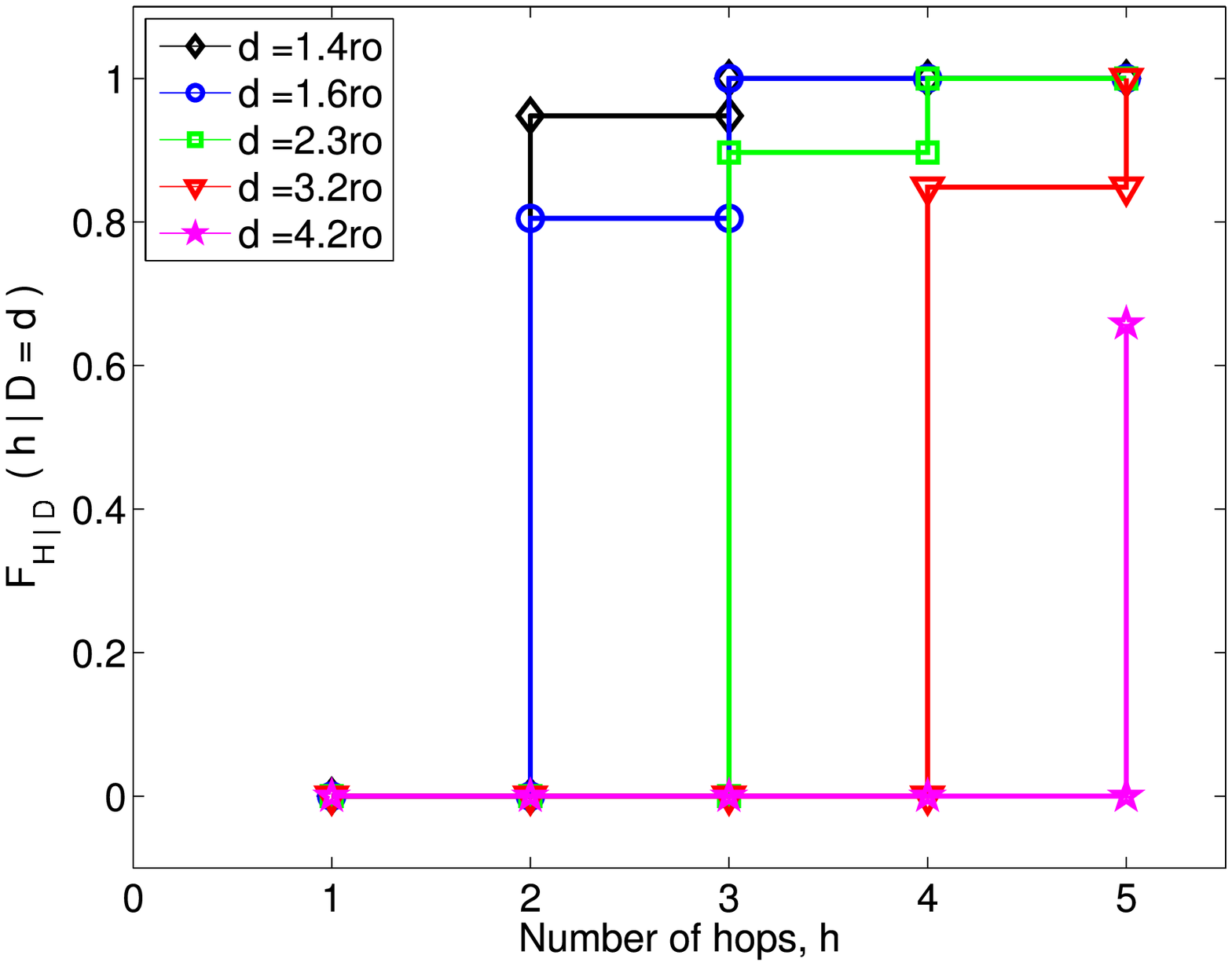}
  \label{fig:hops-given-dist}
    }
  \end{center}
  \caption{Relationship between length of shortest path and Euclidean 
  distance between end points; parameters: $r_0=2, \varrho=1.25$, area:  
  $\mathrm{circle}(R=5\cdot r_0)$}
\end{figure*}

In \cite{Ta2007a}, Ta \etal derive a recursive expression for the
probability $\Phi_k(d)$ that the length of the shortest path between
two nodes is $k$ hops, given their Euclidean distance, $d$. They
assume uniform node distribution and the geometric link model (\cf
\refsec{sec:networks}). From the set $\Phi_k(d)$, $k=1,2,\dotsc$ we may 
derive the discrete conditional probability
distribution function of the shortest path length between two nodes 
given $d$ as
\begin{equation*}
    f_{H|D}(h|D=d) = \sum_{k} {\Phi_k(d)\cdot\delta(h-k)}, h=1,2\dotsc,
\end{equation*}
where $\delta(\cdot)$ is Dirac's delta function. Applying Bayes'
theorem for probability densities (\cf \cite{Durrett:prob}), the 
conditional probability density function
$f_{D|H}(d|H=h)$ of the Euclidean distance between two points
given the number of hops of the shortest connecting path can be
written as
\begin{equation*}
    f_{D|H}(d|H=h) = \frac{\Phi_k(d)\cdot f_D(d)}{\int \Phi_k(d)\cdot f_D(d)}
\end{equation*}
where $f_D(d)$ corresponds to the unconditional probability
distribution function of the distance between two random nodes
over a given surface. This distribution is available in literature
for multiple surfaces, including circles of radius $R$ (``disk
line picking'' problem; \eg \cite{solomon}) and
rectangles of sizes $a$ and $b$ (``rectangle line picking''
problem; \eg \cite{ghosh}).

In \cite{heimlicher2008}, we plot the respective probability and
cumulative distribution functions for multiple combinations of the
transmission range $r_0$ and network density $\varrho$ parameters.
We pick two samples of those plots, \cf \reffig{fig:dist-given-hops} and 
\reffig{fig:hops-given-dist}, to support our arguments.

One may directly observe that for all pairs of curves in the two
plots, $F_{D|H}(d|H=h_1) \le F_{D|H}(d|H=h_0) \; \forall \; h_1
> h_0$ and $F_{H|D}(h|D=d_1) \le F_{H|D}(h|D=d_0) \; \forall \;
d_1 > d_0$. In other words, the stochastic monotonicity
property, as defined in \refdef{def:stoch-monotonicity} pertains to both
distributions, namely the distance is stochastically monotonic in
the path length, $D \mono H$, and vice-versa. The same observation holds 
for all plots in \cite{heimlicher2008}. 

The distribution of the length of the alternate route, \ie the route 
obtained by the intermediate node after the primary route to the destination
breaks, will have a distribution $f_{A|D}(a|D=d)$, which, in the
general case, is different from the one of the primary route. We do
however conjecture the following:
\begin{conjecture} The conditional distribution of the alternate
 route length, $A$,
 given the Euclidean distance, $D$, preserves the stochastic 
 monotonicity property, \ie $A \mono D$.
\end{conjecture}
Given this conjecture, application of \refcor{cor:transitivity} yields 
that the length of the alternate route $A$ is stochastically
monotonic in the length of the primary route $H$, \ie $A \mono H$.


%
\section{Related Work\label{sec:related}}

Route salvaging proposed in the context of MANETs exploits partial paths; failed routes are repaired locally at the point of failure if an alternate route is known (cached). Salvaging might be performed for few packets, as is the case with the Dynamic Source Routing protocol (DSR) \cite{dsr}, or for the whole data stream
\cite{slr}, and has been shown, mainly with simulation studies,
to improve performance. A first step towards analytical comparison of 
end-to-end against local route recovery protocols was presented in 
\cite{Aron00}, yet under a less generic set of assumptions. Finally, 
hop-by-hop transport for wireless networks has also been recently 
explored 
\cite{bibYSHbhbCongestionWirelessMultihop,bibHBMPTransportLayerRevisited}. 
Nevertheless, none of these studies \emph{analytically} addresses the 
existence of partial paths or under which conditions they might help to 
improve performance over traditional end-to-end approaches. In fact, the 
implicit assumption in most of these studies has been that the network 
is almost surely connected.

It was only more recently that this assumption was relaxed in the
context of DTNs~\cite{bibDTNRG}. However, the majority of DTN routing 
protocols
rely on single-hop transfer opportunities (``contacts'') and node 
mobility to eventually deliver a message. In a few cases, hybrid 
approaches that combine elements from both DTN and MANET routing have 
been proposed. In~\cite{Grossglauser:Secon06} the authors look into a 
vehicular network where connected clusters are formed at various 
locations, with little or no connectivity between them. They propose 
regular routing to be used within these clusters, while DTN schemes can 
be used to move messages between clusters when the destination is not 
locally available. Similarly, in~\cite{Ott:DTN-MANET} AODV is used when 
end-to-end connectivity is available, while a node may fall back to DTN 
forwarding when an end-to-end path cannot be discovered. Finally, 
in~\cite{Musolesi:SCAR} a Kalman filtering approach is taken to predict 
the delivery probability (``utility'') of each node in a DTN context, 
but regular routing is used to query and identify the highest utility 
node within the cluster. These approaches more formally target Area 3 of 
the connectivity continuum discussed in \refsec{sec:networks}. Further, 
we believe the analysis presented in this paper could be applied with 
appropriate modifications to such hybrid solutions.

Along a slightly different line of research, there have been
several theoretical studies revisiting the network connectivity
dynamics \cite{Dousse:Thesis,Bett05-shad,Tang2003,hekmat06}.
Whereas the phase transition phenomena for asymptotically large
networks had been already studied in \cite{Philips89,Piret91},
more recent studies have focused on more realistic network scenarios and 
propose metrics that can capture the smoother evolution of connectivity 
with network density for a finite number of nodes. The connectivity 
index (reliability) in \cite{Tang2003} and the giant component in 
\cite{hekmat06}, where a lognormal radio model is superimposed on the 
geometric link model, are two examples of attempts to capture better the 
partially connected nature of finite-size networks and assess the extent 
to which forwarding opportunities are provided by them. The remarks in 
these studies, which are practically in line with our discussion in
\refsec{sec:networks}, have been the main motivation for looking closer  
at the regime of partial connectivity in this paper. Finally, an 
interesting study and classification of the connectivity continuum with 
the goal of understanding what is most suitable for each network class 
is provided in~\cite{Ammar:Chants07}.

\section{Conclusion}\label{sec:conclusion}

In this paper, we have shown that a considerable regime between fully 
connected and fully disconnected networks can be defined, namely 
\emph{partially-connected networks}, where partial paths of multiple 
hops often exist and could be taken advantage of to improve performance. 
DTN schemes usually ignore multi-hop paths altogether and rely on 
single-hop paths (``contacts'') and speculative forwarding. MANET 
schemes, on the other hand, require complete end-to-end paths to 
operate.

We have argued here that local recovery mechanisms (``intermediate 
forwarding''), which store data at the point of path breakage and 
locally try to discover an alternate partial route from that point to 
the destination, can outperform established end-to-end mechanisms 
(``source forwarding'') over a range of connectivity regimes. 
Specifically, we show analytically that \emph{stochastic monotonicity} 
of the length of alternate paths given the length of the current path, 
is a sufficient condition for intermediate forwarding to 
\emph{stochastically dominate} end-to-end forwarding. At the same time, 
we show that, contrary to intuition, weaker conditions relating 
alternate to the current path lengths are not sufficient to prove the 
desired result of superiority.

In future work, we are planning to explore the applicability of our analytical methodology in connectivity regimes where connectivity is close to zero (i.e. end-to-end paths never exist), but reachability is still considerable (Area 3 in \refsec{sec:networks}). Further, we are planning to look in more detail into the protocol-related issues of intermediate forwarding for partially-connected networks, building on the insight acquired by this analytical study.


\bibliographystyle{IEEEtran}
\bibliography{biblio}

\end{document}